\documentclass[%
 ,secnumarabic%
,amssymb, amsmath,nobibnotes, aps, prl]{revtex4}
\usepackage[dvips]{graphicx}
\usepackage{tabularx}
\usepackage{bm}%
\expandafter\ifx\csname package@font\endcsname\relax\else
 \expandafter\expandafter
 \expandafter\usepackage
 \expandafter\expandafter
 \expandafter{\csname package@font\endcsname}%
\fi

\begin{document}

\title{Geometry in transition: A model of emergent geometry}

\author{ Rodrigo Delgadillo-Blando$^{a,b}$, Denjoe O'Connor$^{a}$,  Badis Ydri$^{c}$}
\affiliation{$^{a}$School of Theoretical Physics, DIAS, Dublin, Ireland.\\
$^{b}$Departamento de F\'{\i}sica, CINVESTAV-IPN, M\'exico DF. M\'exico.\\ 
$^{c}$Institut f\"{u}r Physik, Humboldt-Universit\"{a}t zu Berlin, 
D-12489 Berlin, Germany.}

\begin{abstract}
We study a three matrix model with global $SO(3)$ symmetry containing
at most quartic powers of the matrices. We find an exotic line of discontinuous
transitions with a jump in the entropy, characteristic of a 1st order
transition, yet with divergent critical fluctuations and a divergent
specific heat with critical exponent $\alpha=1/2$. The low temperature
phase is a geometrical one with gauge fields fluctuating on a round sphere. 
As the temperature increased the sphere evaporates 
in a transition to a pure matrix phase with no background geometrical
structure.  Both the geometry and gauge fields are determined
dynamically.  It is not difficult to invent higher dimensional models
with essentially similar phenomenology.  The model presents an
appealing picture of a geometrical phase emerging as the system cools
and suggests a scenario for the emergence of geometry in the early
universe.
\end{abstract}

\maketitle

Our understanding of the fundamental laws of physics has evolved to a very geometrical one.
However, we still have very little insight into the origins of geometry itself.
This situation has been undergoing a significant evolution in recent years
and it now seems possible to understand classical geometry as an emergent concept.
The notion of geometry as an emergent concept is not new, see 
for example \cite{Bombelli:1987aa} for an inspiring discussion and 
\cite{Seiberg:2006wf,Ambjorn:2006hu} for 
some recent ideas. We examine such a phenomenon in the context of 
noncommutative geometry \cite{connes} emerging from matrix models,
by studying a surprisingly rich three matrix model
\cite{Azuma:2004zq,CastroVillarreal:2004vh,O'Connor:2006wv}.
The matrix geometry that emerges here has received attention 
as an alternative setting for the regularization of field theories
\cite{Ydri:2001pv,O'Connor:2003aj,Balachandran:2005ew,Grosse:1996mz} and as 
the configurations of $D0$ branes in string theory \cite{Myers:1999ps,Alekseev:2000fd}. Here, however, 
the geometry emerges as the system cools, much as a Bose condensate or superfluid
emerges as a collective phenomenon at low temperatures. There is no background 
geometry in the high temperature phase. 
The simplicity of the model, in this study, allows for a detailed examination 
of such an exotic transition. We suspect the asymmetrical nature of the
transition may be generic to this phenomenon.

We consider the most general single trace Euclidean action (or energy) functional for a 
three matrix model invariant under global $SO(3)$ transformations containing no higher than 
the fourth power of the matrices. This model is surprisingly rich and 
in the infinite matrix limit can exhibit many phases as the parameters are tuned. We find that 
generically the model has two clearly distinct phases, one geometrical the other a matrix phase.
Small fluctuations in the geometrical phase are those of a Yang-Mills
theory and a scalar field around 
a ground  state corresponding to a round two-sphere.
In the matrix phase there is no background spacetime geometry and the fluctuations
are those of the matrix entries around zero. In this note we focus on the subset 
of parameter space where, in the large matrix limit, the gauge group is Abelian.

For finite but large $N$, at low temperature,
the model exhibits fluctuations around a fuzzy sphere \cite{HoppeMadore}.  
In the infinite $N$ limit the macroscopic geometry becomes classical.
As the temperature is increased it undergoes a transition where the entropy jumps, 
yet the model has critical fluctuations and a divergent specific heat. As this critical 
coupling is approached the fuzzy sphere radius expands to a critical radius and the 
sphere evaporates.  The
neighbourhood of the critical point exhibits all the standard symptoms
of a continuous 2nd order transition, such as large scale
fluctuations, critical slowing down (of the Monte Carlo routine) and
is characterized by a specific heat exponent which we argue is
$\alpha=1/2$, a value consistent with our numerical simulations.  In
the high temperature (strong coupling) phase the model is essentially
a zero dimensional Yang-Mills theory.

By studying the eigenvalues of operators in the theory we establish that, in the matrix phase,
the matrices $D_a$ are characterised by continuous eigenvalue distributions which undergo a 
transition to a point spectrum characteristic of the fuzzy sphere phase. The point spectrum is 
consistent with $D_a=L_a/R$ where $L_a$ are $su(2)$ angular momentum generators 
in the irreducible representation given by the matrix size and $R$ is the 
radius of the fuzzy sphere. The full model received an initial study in 
\cite{O'Connor:2006wv} 
while a simpler version invariant under $D_a\rightarrow D_a+\Lambda {\bf 1}$ 
arises naturally as the configuration of $D0$ branes in the large $k$ limit of a boundary 
Wess-Zumino-Novikov-Witten model \cite{Alekseev:2000fd} and has been studied numerically in 
\cite{Azuma:2004zq}. The interpretation here is novel, as are the results on the 
entropy and critical behaviour and the extension to the full model.

Let $D_a$ , $a=1,2,3$, be three $N{\times}N$
Hermitian matrices and let us consider the action
\begin{eqnarray}
\label{main2} 
S&=&S_0+V\ ,
\\
S_0&=&\frac{\tilde{\alpha}^4}{N}\bigg[-\frac{1}{4}Tr[D_a,D_b]^2
+\frac{2i}{3}{\epsilon}_{abc}TrD_aD_bD_c\bigg]
\nonumber
\\
V&=&\frac{m^2\tilde{\alpha}^4}{N}\bigg[- TrD_a^2+\frac{1}{2c_2}
Tr(D_a^2)^2\bigg]\ ,
\nonumber
\end{eqnarray}
so that $\tilde{\alpha}^4=\beta$ plays the r\^{o}le of inverse temperature and we
fix $c_2=\sum_aL_a^2=(N^2-1)/4$ in this study \footnote{By tuning
$c_2$ the resultant gauge group of the Yang-Mills fluctuations can be changed from $U(1)$
while retaining the characteristic features for the phase transition.}.
The absolute minimum of the action is given by $D_a=L_a$. 
Expanding around this configuration, with $D_a=L_a+A_a$, yields a noncommutative Yang-Mills 
action with field strength
\begin{equation}
F_{ab}=i[L_a,A_b]-i[L_b,A_a]+{\epsilon}_{abc}A_c
+i[A_a,A_b]
\end{equation}
and gauge coupling $g^2=1/\tilde{\alpha}^4$.
As written the gauge field includes a scalar field,
\begin{equation}
\Phi=\frac{1}{2}(x_aA_a+A_ax_a+\frac{A_a^2}{\sqrt{c_2}}) \ ,
\end{equation}
as the component of the gauge field normal to the sphere when viewed as embedded in ${\bf R}^3$
where $x_a=L_a/\sqrt{c_2}$.
 
In the large $N$ limit, taken with $\tilde{\alpha}$ and $m$ held fixed
the action for small fluctuations becomes that of a $U(1)$ gauge field coupled to a
scalar field defined on a background commutative two-sphere \cite{CastroVillarreal:2004vh}. 
For large $m$ the
scalar field is not excited. The model with $m=0$ appears as a low energy limit of string 
theory \cite{Alekseev:2000fd}.

One can see the background geometry as that of a fuzzy sphere \cite{HoppeMadore} by noting that 
the $x_a$ satisfy 
\begin{eqnarray}
x_1^2+x_2^2+x_3^2=1~,~
[x_a,x_b]=\frac{i}{\sqrt{c_2}}{\epsilon}_{abc}x_c,
\end{eqnarray}
and the algebra generated by products of the $x_a$ is the algebra of all $N\times N$ 
matrices with complex coefficients. The geometry enters through 
the Laplacian \cite{O'Connor:2003aj} 
\begin{eqnarray}
\hat{\cal L}^2{\bf\cdot}=[L_a,[L_a,{\bf\cdot}]]\ ,
\end{eqnarray} 
which has the same spectrum as the round Laplacian on the commutative sphere,
but cut off at a maximum angular momentum $L=N-1$. The fluctuations
of the scalar have this Laplacian as kinetic term.

The ground state is found by considering the configuration $D_a={\phi}L_a$ where
$\phi$ plays the role of the inverse radius of the sphere. 
The effective potential \cite{CastroVillarreal:2004vh,Azuma:2004ie} 
obtained by integrating out fluctuations 
around this background is given,
in the large $N$ limit, by
\begin{eqnarray}
\frac{V_{\rm eff}}{2c_2}=\tilde{\alpha}^4 \bigg[\frac{{\phi}^4}{4}-\frac{{\phi}^3}{3}
+m^2(\frac{{\phi}^4}{4}-\frac{\phi^2}{2})\bigg]+\log{\phi^2}
\label{V_eff}
\end{eqnarray}
It is unbounded from below at the origin, but, for sufficiently large $\tilde{\alpha}$,
exhibits a local minimum near $\phi=1$ which disappears for $\tilde{\alpha} < \tilde{\alpha}_*$.
The critical curve $\tilde{\alpha}_*$ is determined from the point at which  
the real roots of $\partial V_{\rm eff}/ \partial \phi =0$ merge and disappear.
This occurs at 
\begin{equation}
\phi_*=\frac{\tilde{\phi}_*}{1+m^2}=\frac{3}{8(1+m^2)}\left[1+\sqrt{1+\frac{32t}{9}}\right]
\end{equation}
and gives the critical curve
\begin{equation}
2(1+m^2)^3\frac{{(8/3)}^3}{\tilde{\alpha}^4_*}=
1 + \frac{16t}{3} (1 + \frac{8t}{9}) + (1 + \frac{32t}{9})^{3/2}
\label{pre2}
\end{equation}
where $t=m^2(1 + m^2)$.
This expression interpolates between $\tilde{\alpha}_*={(8/3)}^{3/4}$ with 
$\phi_*=3/4$ for $m=0$ and $\tilde{\alpha}_*={(8/m^2)}^{1/4}$, with 
$\phi_*=1/\sqrt{2}$ for large $m$. 
Thus, as the system is heated, the radius, $R=\phi^{-1}$, expands 
form $R=1$, at large $\tilde{\alpha}$ to $R_*^{-1}=\phi_*$ at $\tilde{\alpha}_*$.
When $\tilde{\alpha}<\tilde{\alpha}_*$ the fuzzy sphere solution no longer exists,
and, in effect, the fuzzy sphere evaporates as the radius approaches $R_*$.

Furthermore, defining 
${\cal S}=<S>/N^2$, we have 
\begin{equation}
{\cal S}=\frac{3}{4}+\frac{i \tilde{\alpha}^4}{12N^2}\epsilon_{abc}Tr(D_aD_bD_c)
-\frac{m^2\tilde{\alpha}^4}{2N^2}Tr(D_aD_a)
\end{equation}
which for the fuzzy sphere phase becomes 
\begin{equation}
{\cal S}=\frac{3}{4}-\tilde{\alpha}^4\left(\frac{1}{24}\phi^3
-\frac{m^2}{8}\phi^2\right)
\end{equation}
with $\phi$ given by the local minimum of $V_{eff}$. 
Expanding in the neighbourhood of the critical point gives
\begin{equation}
{\cal S}={\cal S}_*-\frac{1}{4}\frac{(\tilde{\phi}^2_*+2t\tilde{\phi}_*)}{\sqrt{3\tilde{\phi}_*+4t}}
{(\beta-\beta_c)}^{1/2}+\dots
\label{critentropy}
\end{equation}
and predicts that the transition has a divergent 
specific heat with exponent $\alpha=1/2$. For $m=0$ we find
${\cal S}_*=5/12$ while in the matrix phase it takes the constant value 
${\cal S}=3/4$ hence $\Delta{\cal S}=1/3$,
i.e. the entropy \footnote{For the entropy, ${\cal I}=N^2 {\cal S}+\ln Z$,  the contribution 
from $\ln{Z}$ is continuous and ${\cal S}= \beta U /N^2$, with $U$ the internal energy,
contains the full discontinuity.} 
has a discrete jump of $1/9$ per degree of freedom 
(there are 3 matrices) across the transition. These fractions 
are in excellent agreement with the values obtained 
numerically suggesting that a more complete exact solution 
to the model, at least with $m=0$, may exist.

{\bf Monte Carlo Simulations:}
The model (\ref{main2}) with $m=0$ was studied in \cite{Azuma:2004zq} and an initial study of 
the full model was reported in \cite{O'Connor:2006wv}. In Monte Carlo simulations we use the Metropolis
algorithm and the action (\ref{main2}).
The errors were estimated using the  jackknife method.  

We observe that ${\cal S}$ collapses well for different $N$. Defining 
$\tilde{\alpha}_s$ as the value of $\tilde{\alpha}$ 
at which curves of the average value of the action, $<S>$, for different $N$ cross,
gives a good estimate of the location of the transition.

\begin{figure}
\begin{center}
\includegraphics[width=9cm,angle=-90]{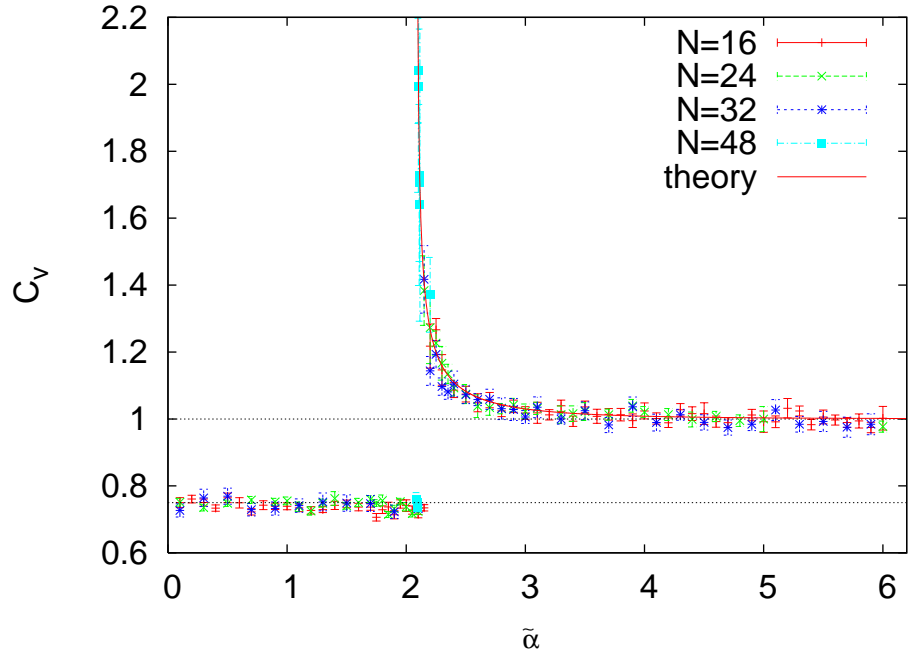}
\caption{{The specific heat for the model $S_0$ as a function of coupling
or inverse temperature $\beta=\tilde{\alpha}^4$. The theoretical curve is obtained from the local minimum of
the effective potential Eq. (\ref{V_eff})}}.
\label{Cvm0}
\end{center}
\end{figure}

For $m=0$ we observe (figure \ref{Cvm0}) a divergence in the specific heat,
$C_v:=<(S-<S>)^2>/N^2$, as the critical value $\tilde{\alpha}_{s}$
is approached from above, i.e. from the low temperature phase. 
For $\tilde{\alpha} > \tilde{\alpha}_{s}$ the model is in the fuzzy sphere 
phase and $C_v$ rapidly approaches 
${C_{v}}=1$ as $\tilde{\alpha}$ is increased. 
For $\tilde{\alpha} < \tilde{\alpha}_{s}$ the model is in a matrix phase and
${C_{v}}=0.75$, retaining this constant value right from the transition at $\tilde{\alpha}_{s}$
to $\tilde{\alpha}=0$. For large $N$ we observe that the location $\tilde{\alpha}_{max}$ 
of the peak in $C_v$ and the minimum $\tilde{\alpha}_{min}$ coincide to the numerical accuracy
we have explored and agree well
with the coupling, $\tilde\alpha_s$, at which $<S>$ intersect for different
values of $N$. The critical coupling determined either as
$\tilde{\alpha}_{max}$ or $\tilde{\alpha}_{min}$ from $C_v$, or as $\tilde{\alpha}_{s}$ 
gives good agreement with (\ref{pre2}) and the 
numerical value found in \cite{Azuma:2004zq}. 

The behaviour when $m\neq0$ is qualitatively similar to $m=0$.
We observe again that, for large $\tilde{\alpha}$, the model is in the fuzzy 
sphere phase where $C_v=1$.

\begin{figure}
\begin{center}
\includegraphics[width=9cm,angle=-90]{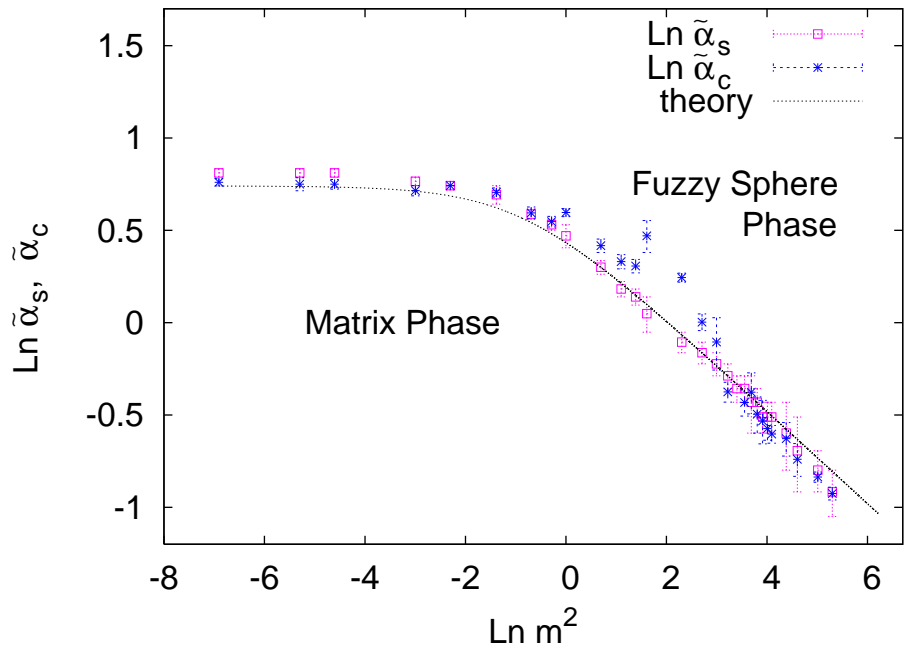}

\caption{{The phase diagram. The theoretical curve is given by Eq. (\ref{pre2}). The specific 
heat exhibits a divergence for small $m$, with the transition becoming smooth 
around $\ln m^2\sim2$ where the curves $\tilde{\alpha}_s$ and $\tilde{\alpha}_c$ separate.}} 
\label{phasediag}
\end{center}
\end{figure}

As $\tilde{\alpha}$ is decreased for fixed $N$, $C_v$ goes 
through a peak at $\tilde{\alpha}_{max}$, with the peak value decreasing as $m^2$ is 
increased. As $\tilde{\alpha}$ is further decreased from $\tilde{\alpha}_{max}$, $C_v$ 
goes through a minimum at $\tilde{\alpha}_{min}$ 
and then increases again, while asymptoting to the value $0.75$ at
$\tilde{\alpha}=0$.
As the value of $m^2$ is increased, our numerical study confirms that 
the fuzzy sphere-matrix model transition is shifted to
lower values of $\tilde{\alpha}$. Extrapolating $m^2\rightarrow\infty$ 
we infer that the critical coupling goes to zero.
This limiting model should be related to
the fuzzy Yang-Mills model without scalar field \cite{Steinacker:2007iq}.
By extrapolating the measured values of $\tilde{\alpha}_{max}$ and $\tilde{\alpha}_{min}$ 
to $N=\infty$ we obtain the critical value $\tilde{\alpha}_{c}$. 
This agrees with $\tilde{\alpha}_s$ and hence we infer both are detecting the matrix-to-sphere phase
transition. However, $\tilde{\alpha}_{c}$ deviates measurably 
from (\ref{pre2}); e.g., for $m^2=200$ $\tilde{\alpha}_{c}= 0.403 \pm 0.014$ while 
(\ref{pre2}) gives $\tilde{\alpha}_*=0.446982$. Thus, though (\ref{pre2}) 
gives a good indicative value for the transition it is not precise.
Our results are summarised in a phase diagram in figure \ref{phasediag}.

We further observe that the jump in entropy decreases as $m$ is increased 
and the nature of the transition seems to change. Our theoretical 
analysis (\ref{critentropy}) still indicates a divergent specific heat 
with exponent $\alpha=1/2$ but with a narrowing critical regime as $m$ is increased.
However, we find no direct evidence for such a regime in the numerical data for large $m$.
We have not been able to determine with any precision where the transition becomes continuous; 
however, the entropy jump seems to disappear at $m^2\sim20$. Also the data for 
$\tilde{\alpha}_s$ and $\tilde{\alpha}_c$ 
separate in this region (see figure \ref{phasediag}) indicating a possible multi-critical point.
It is also possible that beyond this point the transition is in fact 3rd order, a behaviour typical 
of many matrix models; however, the persistence of the 
critical line as determined by $\tilde{\alpha}_s$ the crossing point of the $<S>$ curves
suggests the transition is 2nd order continuous, consistent with the theoretical analysis. 
We know of no other model that exhibits transitions of the type presented here and further 
numerical and theoretical study is warranted.

\begin{figure}[htbp!]
\begin{center}
\includegraphics[width=9cm,angle=-90]{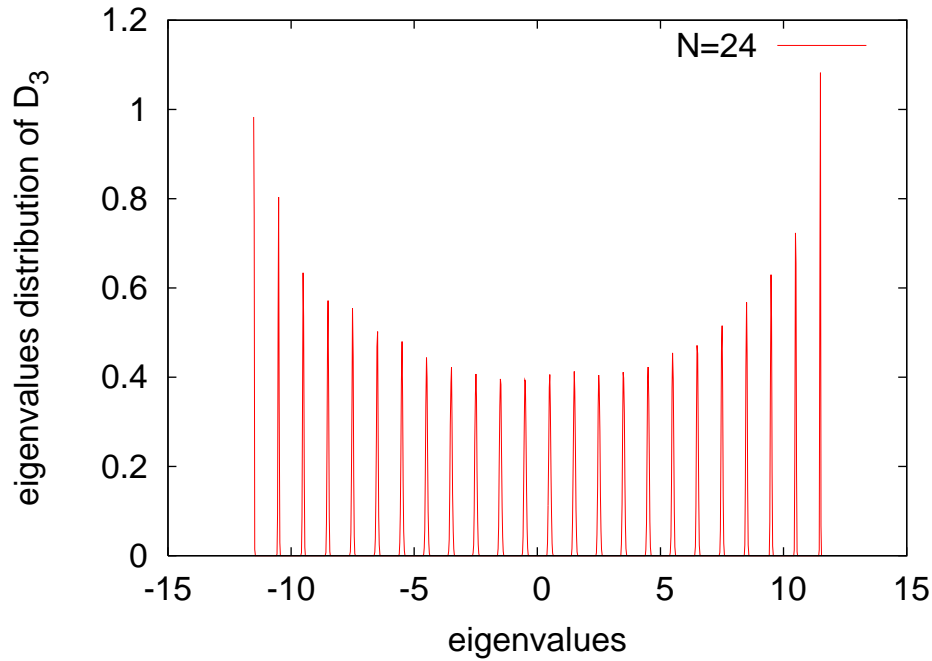}
\caption{The eigenvalue distribution for  
$D_3$, with $N=24$, in the fuzzy sphere for $\tilde{\alpha}=5$ and $m^2=200$. 
It corresponds to $D_3=L_3$. }
\label{D3fuzzy}
\end{center}
\end{figure}

\begin{figure}[htbp!]
\begin{center}
\includegraphics[width=9cm,angle=-90]{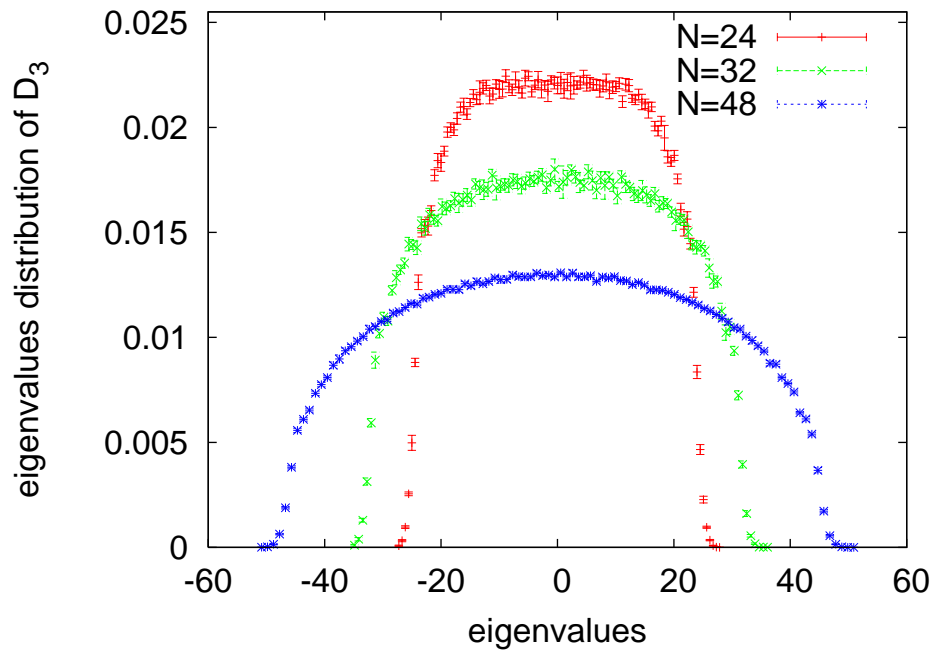}
\caption{The eigenvalue distributions for $D_3$ in the 
matrix phase for $N=24, 32, 48$, $\tilde{\alpha}=0.20$ and $m^2=200$.}
\label{D3matrix}
\end{center}
\end{figure}

More detail on the structure of the phases can be obtained from the distribution 
of eigenvalues of observables. Here we focus on $D_3$ (by rotational symmetry all matrices have 
the same eigenvalue distribution). The characteristic behaviour of the distributions of 
eigenvalues in the fuzzy sphere and matrix phases is illustrated in figures \ref{D3fuzzy}
and \ref{D3matrix} respectively.
Thus we see that, as one crosses the critical curve 
in figure \ref{phasediag}, the eigenvalue distribution of $D_3$ undergoes a 
transition from a point spectrum to a continuous distribution
symmetric around zero, the latter can be fit by the one-cut distribution \cite{Di Francesco:1993nw} 
for a single matrix in a quartic potential. 


{\bf Conclusions:} We have found an exotic transition in a simple three matrix model. The nature 
of the transition is very different if approached from the high or low temperature side.
The high temperature phase is described by Yang-Mills in zero dimensions so there is no
background spacetime geometry. The system in this phase is well
approximated by 3 decoupled matrices each with the same quartic
potential. The value $C_v=0.75$ coincides with the specific heat of $3$ independent matrix models with quartic potential
in the high temperature limit and is therefore consistent with this interpretation.

As the system cools a geometrical phase condenses and at 
sufficiently low temperatures the system is described by small fluctuations 
on a two sphere of a $U(1)$ gauge field coupled to a massive scalar field. 
The critical temperature is pushed upwards as the 
scalar field mass is increased (see figure \ref{phasediag}). We also find evidence that the 
UV/IR mixing \cite{Minwalla:1999px} typical of many non-commutative field theories is linked to the 
geometrical instability and disappears as the scalar mass is increased. 
Once the geometrical phase is well established the specific heat 
takes the value $1$ with the gauge and scalar fields each contributing $1/2$.

The model of emergent geometry described here, though reminiscent of the random matrix approach 
to two dimensional gravity \cite{Ambjorn:2006hu} is in fact very different. 
The manner in which spacetime emerges is
also different from that envisaged in string pictures where continuous eigenvalue 
distributions \cite{Seiberg:2006wf} or a Liouville mode \cite{Knizhnik:1988ak} give rise to extra dimensions. 
It is closely connected to the $D0$ brane scenario described in
\cite{Myers:1999ps} and our model with $m=0$ 
is a dimensionally reduced version of a boundary WZNW models in the
large $k$ limit \cite{Alekseev:2000fd}. It is not difficult to invent higher dimensional models 
with essentially similar phenomenology to that presented here 
(see \cite{Dou:2007in}, \cite{Steinacker:2007dq} and \cite{Kawahara:2007nw}). 

We believe that the model presented above gives an appealing picture of how a geometrical phase might emerge 
as the system cools and suggests a very novel scenario for the emergence of geometry in the 
early universe. In such a scenario the temperature can be viewed as an effect of other degrees of 
freedom present in a more realistic model but not directly participating in the transition we describe.

\paragraph{Acknowledgements}
B.Y. is supported by Marie Curie Fellowship No.MIF1-CT-2006-021797. 
R.D.B. was supported by CONACYT M\'exico and thanks Humboldt-Universit\"{a}t zu
Berlin for hospitality where some computations were performed on HLRN clusters. D.O'C. was partly supported by
EU-NCG Marie Curie Network No. MTRN-CT-2006-031962. We thank Wolfgang Bietenholz for helpful comments.

\end{document}